\documentclass[12pt]{iopart}
\usepackage{iopams}
\usepackage[dvips]{epsfig}
\usepackage{color}
\definecolor{light-gray}{gray}{0.65}
\begin{document}
%23March 2009
\title[Laguerre-type derivatives: Dobi\'nski relations and combinatorial identities]
{Laguerre-type derivatives: Dobi\'nski relations and combinatorial identities}

\author{K.A.Penson$^{a}$, P.Blasiak$^{b}$, A.Horzela$^{b}$,
A.I.Solomon$^{a,c}$ \\
and G.H.E.Duchamp$^{d}$\vspace{2mm}}

\address
{$^a$ Laboratoire de Physique Th\'eorique de la Mati\`{e}re Condens\'{e}e (LPTMC)\\
Universit\'e Pierre et Marie Curie, CNRS UMR 7600\\
Tour 24 - 2i\`{e}me \'et., 4 pl. Jussieu, F 75252 Paris Cedex 05, France\vspace{2mm}}

\address
{$^b$ H. Niewodnicza\'nski Institute of
Nuclear Physics, Polish Academy of Sciences\\
ul. Eliasza-Radzikowskiego 152,  PL 31342 Krak\'ow, Poland\vspace{2mm}}

\address
{$^c$ Physics and Astronomy Department\\
The Open University, Milton Keynes, MK7 6AA, UK}

\address
{$^d$ Institut Galil\'ee, LIPN, Universit\'e Paris-Nord, CNRS UMR 7030\\
99 Av. J.-B. Clement, F 93430 Villetaneuse, France\vspace{2mm}}

\eads{\linebreak \mailto{penson@lptl.jussieu.fr},
\mailto{pawel.blasiak@ifj.edu.pl},
\mailto{andrzej.horzela@ifj.edu.pl},
\mailto{a.i.solomon@open.ac.uk}
\mailto{ghed@lipn-univ.paris13.fr}}


\begin{abstract}
\\
We consider properties of the operators  $D(r,M)=a^r(a^\dag a)^M$ (which we call generalized Laguerre-type derivatives), with $r=1,2,...$ , $M=0,1,...$ ,
where $a$ and $a^\dag$ are boson annihilation and creation operators respectively, satisfying $[a,a^\dag]=1$.
We obtain explicit formulas for the normally ordered form of arbitrary Taylor-expandable functions of $D(r,M)$
with the help of an operator relation which generalizes the Dobi\'nski formula.
Coherent state expectation values of certain operator functions of $D(r,M)$ turn out to be
generating functions of combinatorial numbers. In many cases the corresponding combinatorial structures can be explicitly identified.
\end{abstract}

\maketitle
%%%%%%%%%%%%%%%%%%%%%%%%%%%%%%%%%%%%%%%%%%%%%%%%%%%%%%%%%%%%%%%%%%%%%%%%%%%%%%%%%%%%%%

\section{Introduction}

Among many ways of generalizing the ordinary derivative $\frac{d}{dx}$,
the notion of the so-called Laguerre derivative \cite{Dattoli1} seems to be particularly fruitful.
The idea is to extend the operator $\frac{d}{dx}$ to a simple homogeneous counterpart $D_x$,
which we define as in \cite{Dattoli2},\cite{Dattoli3}
(note that here we omit the factor $(-1)$ present in these references):
\begin{eqnarray}
D_x=\frac{d}{dx}x\frac{d}{dx}.
\end{eqnarray}
In Refs.\cite{Dattoli1},\cite{Dattoli2},\cite{Dattoli3} many important consequences of the replacement $\frac{d}{dx}\rightarrow D_x$
in the integral transform methods and in the operational calculus were investigated.
The link between $D_x$ and the Laguerre polynomials becomes clear if one notices the operational
relation (see Eq.(5) of Ref.\cite{Dattoli3}) which is easy to get using the amusing identity $D_x^{n}=\left(\frac{d}{dx}\right)^{n}x^{n}\left(\frac{d}{dx}\right)^{n}$ \cite{turbiner}
\begin{eqnarray}
\label{ELaguerre}
e^{yD_x}x^n=n!\;y^nL_n(-\frac{\displaystyle x}{\displaystyle y}),
\end{eqnarray}
where $L_n(z)$ are Laguerre polynomials. This  justifies \textit{a posteriori}  the name Laguerre derivative for $D_x$.
Using Eq.~(\ref{ELaguerre}) we may  obtain the  action of $e^{\lambda D_x}$ on various functions, using the different generating functions of Laguerre polynomials listed in Section 5.11 of \cite{Prudnikov}. In particular, using the well known ordinary generating function of $L_n(x)$ (formula 5.11.2.1 for $\alpha=0$ of  \cite{Prudnikov} ) one obtains \cite{Dattoli5}
\begin{eqnarray}
\label{2}
e^{\lambda D_x}e^{-bx}=\frac{1}{1+b\lambda}\exp\left(-\frac{bx}{1+b\lambda}\right)
\end{eqnarray}
valid for $|b\lambda|<1$ \cite{Dattoli5}. Analogously, using the formula 5.11.2.6 of \cite{Prudnikov}, one gets
\begin{eqnarray}
e^{\lambda D_x}{_{1}F_{1}}\left([b],[1],x\right)=\frac{1}{(1-\lambda)^{b}} \ {_{1}F_{1}}\left([b],[1],\frac{x}{1-\lambda}\right),
\end{eqnarray}
with $_{1}F_{1}$ the hypergeometric function\footnote{We use a convenient and self-explanatory notation for the hypergeometric functions of type ${_p}F_{q}$: ${_p}F_{q}$([List of $p$ upper parameters],[List of $q$ lower parameters],$x$).} which for many values of $b$ specializes to elementary or known special functions. Note that for both these examples the action of $e^{\lambda D_x}$ results in a substitution and a prefactor which is reminiscent of the so-called Sheffer-type operators \cite{PLA1}.

We now employ the  operational equivalence
\begin{eqnarray}
\label{aa}
[\frac{d}{dx},x]=1\longleftrightarrow[a,a^\dag]=1,
\end{eqnarray}
where $a$, $a^\dag$
are boson annihilation and creation operators respectively and rewrite $D_x\longleftrightarrow D$ as
\begin{eqnarray}
\label{aaa}
D=aa^\dag a.
\end{eqnarray}
By going one step further we extend Eq.(\ref{aaa}) by defining the generalized Laguerre derivative $D(r,M)$ as
\begin{equation}
\label{araM}
\begin{array}{l}
\hskip -52pt
 {\displaystyle D(r,M)=a^r(a^\dag a)^M \sim D_{x}(r,M) =(\frac{d}{dx})^r(x\frac{d}{dx})^M\ ,
\ \ r=1,2,...\ \ ,M=0,1,...\ .}
\end{array}
\end{equation}
These operators are the object of our present study. Although the equivalence in Eq.(\ref{araM}) between $D(r,M)$ and $D_{x}(r,M)$ is formal since the domains of $a$, $a^\dag$ and $\frac{d}{dx}$, $x$ are different, we shall show that  it provides one with an effective calculational tool.

Since $(a^\dag a)^M$ conserves the number of bosons, the operators $D(r,M)$ act as monomials in boson operators which annihilate $r$ bosons. Recent experiments in quantum optics have shown how one may produce quantum states with specified numbers of photons. This in turn raises the interesting possibility of producing exotic coherent states; that is,  states other than the standard ones which satisfy $a|z\rangle = z|z\rangle$
\cite{Klauder}. The current work introduces operators whose eigenstates may be used to model new coherent states which have many of the features of the standard ones, and  still permit explicit analytic description. The explicit forms of these new generalized coherent states can be used to evaluate relevant physical parameters, such as the photon distribution and the Mandel parameter, squeezing factors and signal-to-noise ratio, etc.

Much theoretical work has been devoted to the description of nonstandard coherent states; for example, the so-called nonlinear coherent states \cite{Vogel}, multiphoton coherent states \cite{Solomon1} and $q$-deformed coherent states \cite{Solomon2}. The structure embodied in definition Eq.(\ref{araM}) is a special case of the extension of boson operators proposed in the construction of nonlinear coherent states \cite{Vogel}. In this latter reference one defines the generalized boson annihilator $b$ by
\begin{eqnarray}
\label{noncoh}
b=af(a^{\dag}a)
\end{eqnarray}
choosing the $f(x)$ that most suits the problem in question. Evidently for this identification $r=1$ and $f(x)=x^M$. In this case the commutator is equal to
\begin{eqnarray}\label{D1D1}
[D(1,M),D^\dag(1,M)]=(a^\dag a+1)^{2M+1}-(a^\dag a)^{2M+1}.
\end{eqnarray}
This emphasizes the fact that although $D(1,M)$ and $D^{\dag}(1,M)$ annihilate and create one boson, respectively, they are \textit{not} canonical boson operators (unless $M=0$). Eq.(\ref{D1D1}) is a special case of
\begin{eqnarray}
\label{10}
[D(r,M),D^\dag(r,M)]&=&(a^\dag a+r)^{2M}\left(\sum_{k=1}^{r+1}|\sigma(r+1,k)|(a^\dag a)^{k-1}\right)-\nonumber\\
&&-
(a^\dag a)^{2M}\left(\sum_{k=1}^{r}|\sigma(r,k)|(a^\dag a)^{k}\right),
\end{eqnarray}
where the $\sigma(r,k)$ are Stirling numbers of the first kind \cite{Comtet}.

Eq.(\ref{10}) was obtained by using the  following two equations
\begin{equation}
\label{11}
\begin{array}{rcl}
a^{r}(a^{\dag})^{r} &=& \prod\limits_{p=1}^{r}{(a^{\dag}a + p)}=\\
&=&\sum\limits_{k=1}^{r+1}|\sigma(r+1,k)|(a^{\dag}a)^{k-1}.
\end{array}
\end{equation}
The first part of Eq.(\ref{11}) is readily proved by induction. To prove the second part of Eq.(\ref{11}) we use the generating function for $|\sigma(r+1,k)|$ in the form \cite{Weisstein}
\begin{equation}
\begin{array}{rcl}
\sum\limits_{k=1}^{r+1}|\sigma(r+1,k)|x^{r+1-k} &=&\prod\limits_{p=1}^{r}(1 + px)\label{13}
\end{array}
\end{equation}
from which, by substituting $x = 1/n$ and using the first part of Eq.(\ref{11}), the second part of Eq.(\ref{11}) follows.

The basic objective of this work is the investigation of arbitrary powers
of $D(r,M)$ which in turn will allow  one to evaluate Taylor-expandable functions of $D(r,M)$. We achieve our goal following recently developed methods of construction of normally ordered products \cite{PLA}, \cite{AnnComb}, \cite{JMP}, \cite {Bugs}.
As we shall show, results derived in this way have a combinatorial flavour and lend themselves to a
combinatorial interpretation.

The paper is organized as follows.
In Section \ref{NormalOrd} we introduce generalizations of the Stirling and Bell numbers which are well known from classical combinatorics and relate them to the normally ordered powers
of operators $D(r,M)$. These numbers, as shown in Section \ref{Dobinski}, may be explicitly found using generalized Dobi\'nski relations.  In Section \ref{Combinatorics} we compare  calculations of purely analytical origin with those based on methods of graph theory and give a combinatorial interpretation of our results. Examples of various applications of our approach are presented in Section \ref{Examples} while Section \ref{Conclusions} summarizes the paper.

\section{Normal ordering: Generalized Stirling and Bell Numbers}
\label{NormalOrd}

The \emph{normally ordered} form of $F(a,a^\dag)$, denoted by $F_{\mathcal{N}}(a,a^\dag)$
\cite{Louisell} is obtained by moving all annihilators to the right using the canonical commutation
relation of Eq.(\ref{aa}). It satisfies $F_{\mathcal{N}}(a,a^\dag)=F(a,a^\dag)$. On the
other hand the \emph{double dot} operation $:G(a,a^\dag):$ means that we are applying the
same ordering procedure but without taking account of the commutation relation.
Conventionally the solution to the normal ordering problem is obtained if a function
$G(a,a^\dag)$ is found satisfying
\begin{eqnarray}\label{NFG}
F_{\mathcal{N}}(a,a^\dag)=\ :G(a,a^\dag):\ \ .
\end{eqnarray}
A large body of research has been recently devoted to finding the solution of
Eq.(\ref{NFG}) \cite{AmJPhys}. A general approach which facilitates a combinatorial interpretation of
quantum mechanical quantities is to use the coherent state representation. Standard
coherent states
\begin{equation}\label{scs}
  |z\rangle=e^{-|z|^2/2}\sum_{n=0}^\infty\frac{z^n}{\sqrt{n!}}|n\rangle
\end{equation}
 with the  number states $|n\rangle$ satisfying  $a^\dag a|n\rangle=n|n\rangle$, $\langle
n|n'\rangle=\delta_{n,n'}$ and $z$ complex, are eigenstates of the annihilation operator,
i.e. $a|z\rangle=z|z\rangle$. The latter eigenstate property shows that having solved the
normal ordering problem Eq.(\ref{NFG}) for an operator $F(a,a^\dag)$ we immediately find
\begin{eqnarray}
\label{NFG2} \langle z|F_{\mathcal{N}}(a,a^\dag)|z\rangle =\ G(z,z^{*}) \ .
\end{eqnarray}
An early  observation on how to extract combinatorial content from normally ordered forms
\cite{Katriel} was based on the formula $e^{\lambda a^\dag a}=\ :e^{a^\dag
a(e^\lambda-1)}:$ \cite{Cahill}. It led to the identification
\begin{eqnarray} \label{Bell}
\langle z|(a^\dag a)^n|z\rangle\stackrel{z=1}{=}B(n),
\end{eqnarray}
where the $B(n)$ are conventional Bell numbers described in \cite{Comtet}. Eq.(\ref{Bell})
may be taken as a {\em definition} of the Bell numbers.  For Stirling numbers of the
second kind we have \cite{Comtet},
\begin{equation}\label{stirling}
  ( a^\dag a)^n= \sum_{k=0}^{n}S(n,k)(a^\dag)^ka^k
\end{equation}
(which may also be used as a practical definition)  in terms of which one defines the
Bell polynomials by
\begin{equation}\label{Bellpoly}
  B(n,x)=\sum_{k=0}^{n}S(n,k)x^k.
\end{equation}

 We have extended and developed the coherent state
methodology for operators other than $a^\dag a$ in \cite{PLA}, \cite{AnnComb} and \cite{JMP}.

After the seminal observation by Katriel \cite{Katriel}, combinatorial methods found
widespread application in this context \cite{PLA},\cite{AnnComb},\cite {JMP},\cite{Bugs},\cite{Mikhailov}.  We apply these methods to $F(a,a^\dag)=[D(r,M)]^n$, $n=1,2,...$ .

Formally, we write $[D(r,M)]^n$ in  normally
ordered form as
\begin{eqnarray}
\label{Saa} [D(r,M)]^n=\left[\sum_{k=0}^{Mn}S_{r}^{(M)}(n,k)(a^\dag)^ka^k\right]a^{rn}.
\end{eqnarray}

Clearly, from Eq.(\ref{Saa}) the integers $S_{r}^{(M)}(n,k)$ are generalizations of
the conventional Stirling numbers of the second kind (which are recovered for $r=0,
M=1$). Analogously to Eq.(\ref{Bellpoly}) the numbers $S_{r}^{(M)}(n,k)$ serve to define
the generalized Bell polynomials
\begin{eqnarray}
\label{B} B_{r}^{(M)}(n,x)=\sum_{k=0}^{Mn}S_{r}^{(M)}(n,k)x^k.
\end{eqnarray}
Finding the explicit form of these generalized Stirling numbers will give the normally
ordered form of $[D(r,M)]^n$. We proceed to do this in the next section by use of a
generalization of the famous Dobi\'nski formula.

\section{Generalized Dobi\'nski formula}\label{Dobinski}
 We first write  Eq.(\ref{Saa}) in derivative form as
 \begin{eqnarray} \label{Sdd}
[D_x(r,M)]^n=
\left[\sum_{k=0}^{Mn}S_{r}^{(M)}(n,k)x^k\left(\frac{d}{dx}\right)^k\right]\left(\frac{d}{dx}\right)^{rn}.
\end{eqnarray}

Acting with the r.h.s. of Eq.(\ref{Sdd}) on $e^x$ one obtains $B_{r}^{(M)}(n,x)e^x$. The
action of the l.h.s. of Eq.(\ref{Sdd}) on $e^x$ is obtained by acting with generalized
Laguerre derivatives on monomials $x^p$
\begin{eqnarray}
D_x(r,M)\ x^p=p^{\underline{r}}p^Mx^{p-r},
\end{eqnarray}
where $p^{\underline{r}}=p(p-1)...(p-r+1)$ is the falling factorial, then extending it to the $n$-th power
\begin{eqnarray}
\label{Dpxn}
\left[D_x(r,M)\right]^nx^p=\left[\prod_{j=0}^{n-1}(p-rj)^{\underline{r}}(p-rj)^M\right]
x^{n-rp},
\end{eqnarray}
and next summing up contributions for $x^p/p!$
\begin{eqnarray}
\label{Dpxn1}
\sum\limits_{p=0}^{\infty}\left[D_x(r,M)\right]^n{\displaystyle{\frac{x^p}{p!}}}=\sum\limits_{p=rn}^{\infty}\left[\prod_{j=0}^{n-1}(p-rj)^{\underline{r}}(p-rj)^M\right]{\displaystyle{\frac{x^{p-rn}}{p!}}}\ .
\end{eqnarray}
Upon simplifying Eq.(\ref{Dpxn1}) leads to the Dobi\'nski-type
representation of generalized Bell polynomials \cite{Bugs},\cite{Lognormal},\cite{Dobinski}:
\begin{eqnarray}
\label{Dob} B_{r}^{(M)}(n,x)=e^{-x}\sum_{l=0}^\infty
\left[\prod_{i=1}^n(l+ir)\right]^M\frac{x^l}{l!}\ ,
\end{eqnarray}
verified by direct calculation of $\langle z|[a^{r}(a^{\dag}a)^{M}]^{n}|z\rangle$.
The classic  Dobi\'nski formula \cite{Comtet} corresponds to $r=0,M=1$:
 \begin{eqnarray}
\label{classDob} B(n,x)=e^{-x}\sum_{l=0}^\infty \frac{l^n x^l}{l!}.
\end{eqnarray}

From Eq.(\ref{Dob}) the generalized Stirling numbers are obtained by standard Cauchy
multiplication of series
\begin{equation}
\label{SrM}
\begin{array}{l} 
\hskip -25pt {S_{r}^{(M)}(n,k)={\displaystyle\frac{1}{k!}\sum_{j=0}^k\left(\begin{array}{c}{k}\\{j}\end{array}\right)(-1)^{k-j}
\left[\prod_{i=1}^n(j+ir)\right]^M}\ ,
\ \ \ k=0,1,\dots,Mn.}
\end{array}
\end{equation}
We point out that Eqs.(\ref{Dob}) and (\ref{SrM}) are the central results we need for further
calculations. For practical applications it is useful to note that the
generalized Stirling and Bell numbers, as well as generalized Bell polynomials, can be
expressed through generalized hypergeometric functions $_{p}F_{q}$.

 Below we quote some examples of such relations.

\begin{equation}
\label{stgen}
S_{1}^{(M)}(n,k)= \frac{(-1)^k (n!)^M}{k!}\cdot\!\!\ _{M+1}F_{M}([-k,\underbrace{n+1,...,n+1}_{M\ times}],[\underbrace{1,...,1}_{M\ times}],1)
\end{equation}

\begin{equation}
\label{bellgenpol}
B_{1}^{(M)}(n, x)= e^{-x}(n!)^M\cdot\!\!\ _{M}F_{M}([\underbrace{n+1,...,n+1}_{M\ times}],[\underbrace{1,...,1}_{M\ times}],x)
\end{equation}

The numbers $B^{(M)}_{1}(n)= B^{(M)}_{1}(n,1)$  can be shown to be related to the numbers $B_{p,p}(n)$ (introduced in Refs.\cite{PLA} and \cite{AnnComb}) characterizing the normal order of $[(a^{\dag})^{p}a^{p}]^{n}$  by the formula
\begin{equation}
\label{bellgenpol1a}
B^{(M)}_{1}(n)=B_{n,n}(M+1)\ ,
\end{equation}
as seen by comparing Eq.(2.6) in Ref.\cite{AnnComb} with  Eq.(\ref{Dob}) of the present work.
\begin{equation}
\label{bellgenpol1}
\begin{array}{l}
\hskip -78pt{B_{2}^{(M)}(n, x)= \frac{1}{{\pi}^{M/2}}2^{Mn}e^{-x}\left((n!)^M\cdot\!\!\ _{M}F_{M+1}([\underbrace{n+1,...,n+1}_{M\ times}],[\!\underbrace{1,...,1}_{M\ times}\!,1/2],x^2/4){\pi}^{M/2}+\right.}\\
\hskip -78pt{\left.+2^{M}\left[{\Gamma(n+3/2)}\right]^M\cdot\!\!\ x \cdot {_{M}F_{M+1}}([\underbrace{n+3/2,...,n+3/2}_{M\ times}],[\underbrace{3/2,...,3/2}_{M+1\ times}],x^2/4)\right) \ ,}
\end{array}
\end{equation}
\begin{equation}
\hskip -1pt{\label{bellgenpol2}}
\begin{array}{l}
\hskip -78pt{B_{3}^{(M)}(n,x)= \frac{1}{2^{M+1}[\pi\Gamma(2/3)]^{M}}e^{-x}\ \times}
\\
\hskip -78pt{\times\ \left(2^{M+1}3^{Mn}[\pi n! \Gamma(2/3)]^{M}\cdot\!\!\ _{M}F_{M+2}([\underbrace{n+1,...,n+1}_{M\ times}],[\!\underbrace{1,...,1}_{M\ times}\!,1/3,2/3],x^3/27) + \right. }\\\\
\hskip -78pt{\left.+ 2\cdot 3^{M(n+3/2)} \left(\Gamma^{2}(2/3)\Gamma(n+4/3)\right)^M\right.\times} \\\\
\hskip -78pt{\left. \times\  x\cdot {_{M}F_{M+2}}([\underbrace{n + 4/3,...,n + 4/3}_{M\ times}],[\underbrace{4/3,...,4/3}_{M+1\ times},2/3],x^3/27)\  + \right. }\\
\hskip -78pt{\left.+ 3^{M(n+1)}\left[\pi{\Gamma(n+5/3)}\right]^M\cdot x^{2}\cdot\!\!\ {_{M}F_{M+2}}([\underbrace{n+5/3,...,n+5/3}_{M\ times}],[\underbrace{5/3,...,5/3}_{M+1\ times},4/3],x^3/27)\right)\ .}
\end{array}
\end{equation}

We conjecture that in general $B^{(M)}_{r}(n,x)$ is a combination of $r$ hypergeometric functions of type $_{M}F_{M+r-1}$ of argument $x^{r}/{r^{r}}$.

Examples of numbers resulting from Eqs.(\ref{bellgenpol})-(\ref{bellgenpol2}) for $n=0,\dots,6,\dots$ are
\begin{equation}
\label{numbers}
\begin{array}{l}
\hskip -78pt{M=1\ \ \ \ B_{1}^{(1)}(n) = 1,2,7,34,209,1546,13227\dots\ ,}\\\\
\hskip -78pt{M=2\ \ \ \ B_{1}^{(2)}(n) = 1,5,87,2971,163121,12962661\dots\ ,}\\\\
\hskip -78pt{M=3\ \ \ \ B_{1}^{(3)}(n) =  1,15,1657,513559,326922081,363303011071\dots\ ,}\\\\
\hskip -78pt{M=2\ \ \ \ B_{2}^{(2)}(n)  =  1,10,339,23395,2682076,457112571,107943795145\dots\ ,}\\\\
\hskip -78pt{M=3\ \ \ \ B_{2}^{(3)}(n) = 1,37,9415,7063615,11360980081,33040809105661,}\\\\
\hskip -78pt{\ \ \ \ \ \ \ \ \ \ \ \ \ \ \ \ \ \ \ \ \ \ 156151310977544887\dots\ ,}\\\\
\hskip -78pt{M=3\ \ \ \ B_{3}^{(3)}(n) =  1,77,39839,62310039,214107236041,1358185668416501,} \\\\
\hskip -78pt{\ \ \ \ \ \ \ \ \ \ \ \ \ \ \ \ \ \ \ \ \ \ 14247249149298651007\dots\ ,}\\\\
\hskip -78pt{M=4\ \ \ \ B_{3}^{(4)}(n) = 1,372,1905633,43249617004,2805942285116705,}\\\\
\hskip -78pt{\ \ \ \ \ \ \ \ \ \ \ \ \ \ \ \ \ \ \ \ \ \ 411223445534704016116,117428972441699060660584977\dots\ ,}
\end{array}
\end{equation}
which are  positive integers and as such admit combinatorial interpretation. The first two sequences in Eq.(\ref{numbers}) may be identified as A002720 (which enumerates matching numbers of a perfect graph $K(n,n)$) and A069948, respectively, in Ref.~\cite{sloane}.

We note in passing that the numbers $B_{r}^{(M)}(n)$ are solutions of the Stieltjes moment problem, \textit{i.e.} they are the $n$-th moments of positive weight functions on the positive half axis. This can be deduced from their Dobi\'nski-type relations Eq.(\ref{Dob}), whose form allows one to obtain the weight functions for any $r$ and $M$. For the first two sequences in Eq.(\ref{numbers}) the Stieltjes weights are given in \cite{sloane} under their entries.

As a second illustration of our approach we shall apply it to $D(r,1)$. Note that
\begin{equation}
\label{bbb}
\left[a^r(a^\dag a)^M\right]^n=\,:B_r^{(M)}(n,a^\dag a)\,a^{rn}:\ ,
\end{equation}
which upon using the Dobi\'nski relation Eq.(\ref{Dob}) for $M=1$ leads to
\begin{eqnarray}
\label{shef}
e^{\lambda D(r,1)}=\ :\frac{1}{1-\lambda ra^r}\exp\left(\frac{a^\dag a}{(1-\lambda ra^r)^{1/r}}-a^\dag a\right):\ .
\end{eqnarray}
The operator $D(r,1)$
is of Sheffer-type viewed through hermitean conjugation (see Refs. \cite{PLA1},\cite{turbiner2}) and Eq.(\ref{shef}) can also be obtained through the methods developed in Ref.\cite{PLA1} (see Appendix). Consequently,
\begin{eqnarray}
\label{E}
\langle z| e^{\lambda D(r,1)}|z\rangle&\stackrel{z=1}{=}&\frac{1}{1-r\lambda}\exp\left(\frac{1}{(1-r\lambda)^{1/r}}-1\right)\equiv \\
&\equiv&\sum_{n=0}^\infty B_{r}^{(1)}(n)\frac{\lambda^n}{n!},
\end{eqnarray}
where
\begin{eqnarray}
\label{Bsigma}
\langle z|\left(a^ra^\dag a\right)^n|z\rangle\stackrel{z=1}{=}B_{r}^{(1)}(n)=\sum_{p=1}^{n+1}|\sigma(n+1,p)|r^{n-p+1}B(p-1)\ .
\end{eqnarray}
In Eq.(\ref{Bsigma}) $\sigma(n,k)$ are  Stirling numbers of the first kind, $B(n)$ are conventional Bell numbers, and in obtaining Eq.(\ref{Bsigma}) we have again used Eq.(\ref{11}).

Using Eqs.(\ref{Saa}) and (\ref{bellgenpol}) we obtain for $r=1$ the following formula in a compact notation
\begin{eqnarray}
\label{BBB1}
\!\!\!\!\!\!\left[D(1,M)\right]^n=\ (n!)^M :e^{-a^\dag a}\cdot\!\!\ _{M}F_{M}([\underbrace{n+1,...,n+1}_{M\ times}],[\underbrace{1,...,1}_{M\ times}],a^\dag a)a^n:\ .
\end{eqnarray}
The last formula can be used to normally order $H(\lambda D(1,M))$ for any Taylor-expandable $H(x)$.

\section{Combinatorics of normally ordered Laguerre derivatives}\label{Combinatorics}

\begin{figure}[h]
\resizebox{\columnwidth}{!}{\includegraphics{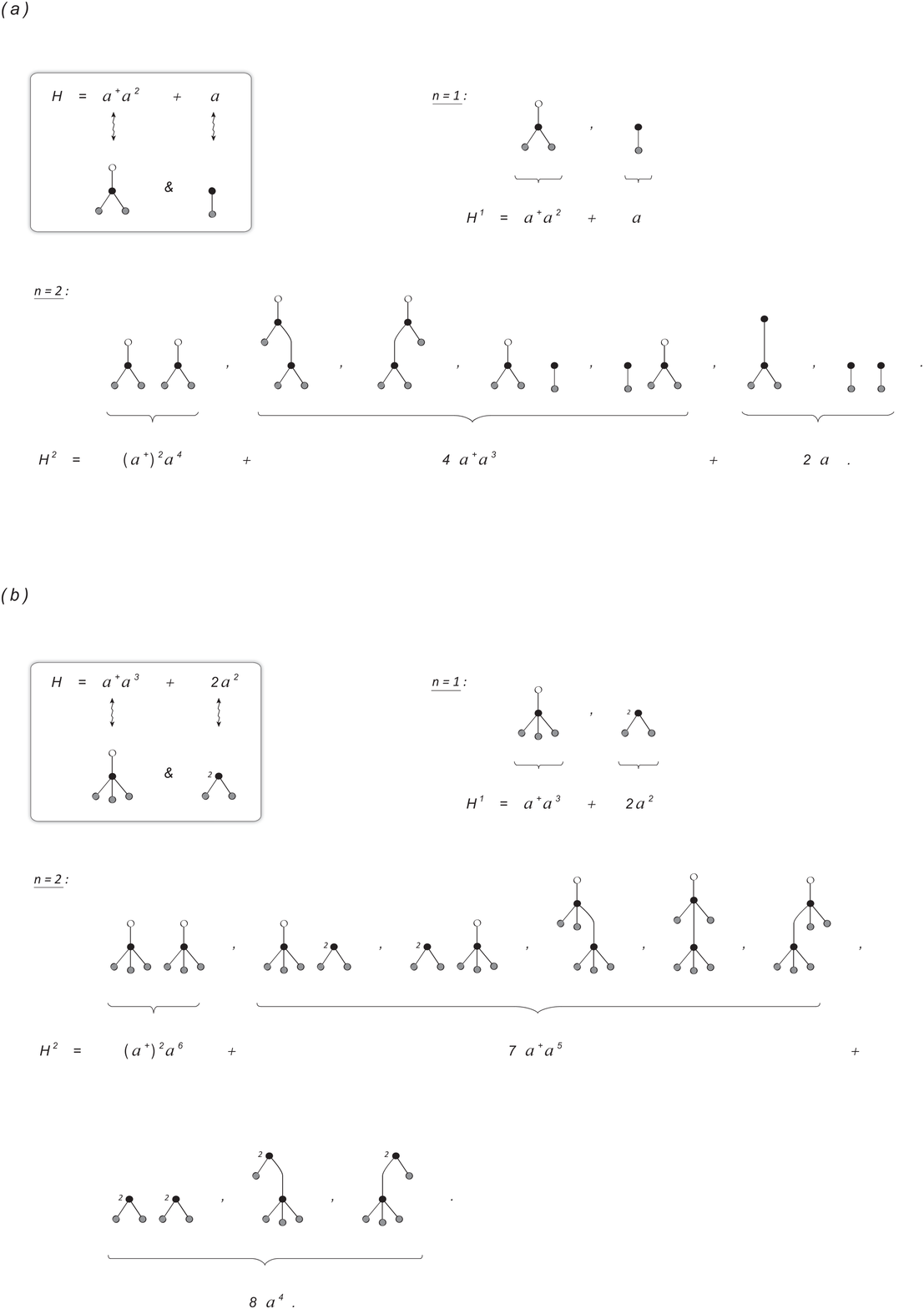}}
\caption{\label{LaguerreFig} Building blocks (in the inset) and the associated graphs of order $n=1,2$ for Laguerre derivatives: (a) $H=D(1,1)=a^{\dag}a^{2}+a$ and (b) $H=D(2,1)=a^{\dag}a^{3}+2a^{2}$.}
\end{figure}

In previous Sections we considered the normal ordering of Laguerre derivatives for which the results heavily exploited combinatorial identities stemming from the underlying iterative character of the problem. Indeed, the reordering of the operators $a$ and $a^{\dag}$ is a purely combinatorial task which can be interpreted in terms of graphs \cite{Bugs},\cite{arXiv},\cite{HW} and analyzed by the use of combinatorial constructors \cite{flajolet}. Briefly, to each operator in the normally ordered form
$H=\sum_{r,s}\alpha_{r,s}\,(a^{\dag})^{r}a^s$ one associates a set of one-vertex graphs such that each vertex $\bullet$ carries  weight $\alpha_{r,s}$ and has $r$ outgoing and $s$ incoming lines whose free ends are marked with white $\circ$ and gray ${\color{light-gray}\bullet}\!\!\!\circ$ spots respectively. Multi-vertex graphs are built in a step-by-step manner by adding one vertex at each consecutive step and joining some of its incoming lines with some the free outgoing lines of the graph constructed in the previous step. Additionally, one keeps track of the history by labeling each vertex by the number of steps in which it was introduced. As a result, one obtains a set of increasingly labeled multi-vertex graphs with some free incoming and outgoing lines. It can be shown that the normal ordering of powers of the operator $H$ can be obtained by enumeration of such structures. Namely, the coefficient of $({a}^{\dag})^{k}a^l$ in the normally ordered form of the operator $H^n$ is obtained by counting all possible graphs with $n$ vertices $\bullet$ and having $k$ white $\circ$ and $l$ gray ${\color{light-gray}\bullet}\!\!\!\circ$ spots respectively. For illustration, we give two examples of Laguerre derivatives $D(1,1)=aa^{\dag}a=a^{\dag}a^2+a$ and $D(2,1)=a^2a^{\dag}a=a^{\dag}a^3+2\,a^2$ and their graph representation leading to the solution of the normal ordering problem by simple enumeration (see Fig.~\ref{LaguerreFig}).
One should compare these ``graphical results'' with the explicit formulas of Eqs.(\ref{SrM}) and (\ref{Dob}) or the expansion coefficients of the generating function in Eq.(\ref{E}) for $r=1,2$ and $M=1$. Thus, using Eq.(\ref{numbers}),  the coefficients multiplying the operators in Fig.(1a) are the first two terms in $B^{(1)}_1(n) = 2,7,34,209$ for $n=1,2,\ldots$ (A002720). Similar coefficients in Fig.(1b) are the first two terms in $B^{(1)}_2(n) = 3,16,121,1179$ for $n=1,2,\ldots$ (A121629).

\section{Examples}\label{Examples}
%Below we quote some examples.

1. For $r=M=1$, i.e. for $D(1,1)=aa^\dag a$\,  one obtains (see \cite{Riordan} for a similar calculation):
\begin{eqnarray}
\label{DL}
\left[D(1,1)\right]^n=n!:L_n(-a^\dag a):a^n,
\end{eqnarray}
where $L_n(y)$ are Laguerre polynomials and Eq.(\ref{DL}) is derived from Eq.(\ref{BBB1}) and using the definition of $L_n(y)$ via the function $\ _1F_1$. Then
\begin{equation}
\label{xxxx}
\begin{array}{rcl}
e^{\lambda D(1,1)}&=&{\displaystyle\sum_{n=0}^\infty \frac{\lambda^n}{n!}\left[{D(1,1)}\right]^n = }\\ 
&=&{\displaystyle\ :\sum_{n=0}^\infty L_n(-a^\dag a)(\lambda a)^n:\ =\ :\frac{1}{1-\lambda a}\exp\left(\frac{\lambda a^\dag a^2}{1-\lambda a}\right):\ ,}
\end{array}
\end{equation}
 where in Eq.(\ref{xxxx}) we have used the ordinary generating function for the Laguerre polynomials \cite{Prudnikov}.

Using other generating functions listed on p.704 of Ref.\cite{Prudnikov} one can
derive further formulas of type Eq.(\ref{xxxx}). ( In a) and b) below: $\lambda\neq0,\ p=1,2,...$).

a) Formula $5.11.2.6$ of \cite{Prudnikov} for $\alpha=0$ provides the normal ordering of
\begin{eqnarray}
\label{example1}
_{1}F_{1}\left([b],[1],\lambda D(1,1)\right) = {\displaystyle :\!\frac{1}{(1-\lambda a)^{b}}{_{1}F_{1}}\left([b],[1],\frac{\lambda a^{\dag} a^{2}}{1-\lambda a}\right)\!:\ ,}
\end{eqnarray}
which for $b$ integer and half-integer can be written down in terms of known functions. Examples are:
\begin{equation}
{\hskip -19pt\label{example2}}
\begin{array}{l}
\hskip -50pt{{_{1}{F}_{1}}\left([3],[1],\lambda D(1,1)\right) = {\displaystyle :\!\frac{1}{(1-\lambda a)^{3}}
L_{2}\left(-\frac{\lambda a^{\dag} a^{2}}{1-\lambda a}\right)\exp{\left(\frac{\lambda a^{\dag} a^{2}}{1-\lambda a}\right)}\!:\ ,}}\\\\
\hskip -50pt {{_{1}{F}_{1}}\left([\frac{3}{2}],[1],\lambda D(1,1)\right) = {\displaystyle :\exp{\left(\frac{\lambda a^{\dag} a^{2}}{2(1-\lambda a)}\right)}\times}}\\\\
\hskip 78pt {\displaystyle{\times\left[I_{0}\left({\frac{\lambda a^{\dag} a^{2}}{2(1-\lambda a)}}\right )(1+\frac{\lambda a^{\dag} a^{2}}{1-\lambda a}) + I_{1}\left({\frac{\lambda a^{\dag} a^{2}}{2(1-\lambda a)}}\right)\right]\!:\ .}}
\end{array}
\end{equation}
where $I_0(y)$ and $I_1(y)$ are modified Bessel functions.

b) Similarly, we consider the formula $5.11.2.8$ of \cite{Prudnikov}:
\begin{eqnarray}
\sum_{n=0}^\infty\left(\begin{array}{c}{n+p}\\{n}\end{array}\right)t^nL_{n+p}(x)=\frac{1}{(1-t)^{p+1}}e^{-\frac{tx}{1-t}}L_p\left({\frac{x}{1-t}}\right)\ .
\end{eqnarray}
Using Eq.(\ref{DL}) we obtain the normally ordered form of $[\lambda D(1,1)]^p\exp(\lambda D(1,1))/p!\,$:
\begin{equation}
\label{example3}
\begin{array}{rcl}
{\displaystyle
\frac{1}{p!}\sum_{n=0}^\infty\frac{[\lambda D(1,1)]^{n+p}}{n!}}&=&{\displaystyle\sum_{n=0}^\infty\frac{(n+p)!}{p!n!}:L_{n+p}(-a^\dag a)(\lambda a)^{n+p}: \ =}\\
&=&{\displaystyle :\frac{1}{(1-\lambda a)^{p+1}}e^{\frac{\lambda a^\dag a^2}{1-\lambda a}}L_p(-\frac{a^\dag a}{1-\lambda a}):\ (\lambda a)^p\ .}
\end{array}
\end{equation}

2. The normal order of the modified Bessel function of the first kind $I_0\left(2(\lambda D(1,1))^{1/2}\right)$ may be derived:
\begin{equation}
\label{example4}
\begin{array}{rcl}
I_0(2(\lambda D(1,1))^{1/2})&\!=\!&\!{\displaystyle :\sum_{n=0}^\infty \frac{L_n(-a^\dag a)(\lambda a)^n}{n!}: =}\\
&\!=\!&\!{\displaystyle :e^{\lambda a}J_0(2\sqrt{\lambda(-aa^\dag a)}):\ =\ :e^{\lambda a}I_0(2\sqrt{\lambda a^\dag a^2}):
\ ,}
\end{array}
\end{equation}
where in the last line we have used the exponential generating function of Laguerre polynomials \cite{Prudnikov}. The analogous formula for $J_0\left(2(\lambda D(1,1))^{1/2}\right)$ reads
\begin{equation}
\label{example5}
\begin{array}{rcl}
J_0(2(\lambda D(1,1))^{1/2})&=&{\displaystyle\ :\sum_{n=0}^\infty \frac{L_n(-a^\dag a)(-\lambda a)^n}{n!}:} =\\
&&{\displaystyle =\ :e^{-\lambda a}I_0(2\sqrt{\lambda a^\dag a^2}):\ .}
\end{array}
\end{equation}

3. We quote here the eigenfunctions of $D_x(r,M)$ with eigenvalue 1 satisfying $D_x(r,M)E(r,M;x)=E(r,M,x)$,
with the following $r$ boundary conditions:
\begin{eqnarray}
E(r,M;0)= 1,\ \ \ \ \ \ \ \left.\left[\frac{d^p}{{dx}^{p}}E(r,M;x)\right]\right|_{x=0}=0,\ \ \ p=1,\dots,r-1
\end{eqnarray}
which are
\begin{eqnarray}\label{EEEEE}
E(r,M;x)=\ _0F_{M+r-1}([\; \;],[\underbrace{1/r,2/r,...,(r-1)/r,}_{{r-1}\  times}\underbrace{1,...,1}_{M\  times}],x^r/r^{r+M})\ .
\end{eqnarray}
Useful normal ordering formulas can be obtained by applying the Dobi\'nski relations
to the eigenfunctions of $D_x(1,M)$ {\em with the argument taking operator values}, see Eq.(\ref{EEEEE}), i.e. $E(1,M;D_x(1,M))$. We briefly show the calculation, in boson notation, for $E(1,2;\lambda D(1,2))=_0\!\!F_2([\; \;],[1,1],\lambda D(1,2))$, see Eq.(\ref{BBB1}) :
\begin{equation}
\label{52}
\begin{array}{l}
_0F_2([\; \;],[1,1],\lambda D(1,2))=\sum\limits_{n=0}^\infty{\displaystyle\frac{\lambda^n}{(n!)^3}}[a(a^\dag a)^2]^n=\\\\
=\ :e^{-a^\dag a}\sum\limits_{l=0}^\infty{\displaystyle{\frac{(a^\dag a)^l}{l!}}}\sum\limits_{n=0}^\infty{\displaystyle\frac{((n+l)!)^2}{(l!)^3}}(\lambda a)^n:\ =\\\\
=\ :e^{-a^\dag a}\sum\limits_{l=0}^\infty{\displaystyle\frac{(a^\dag a)^l}{l!}}\ _2F_2([1+l,1+l],[1,1],\lambda a):\ ,
\end{array}
\end{equation}
and similarly
\begin{equation}
\label{a1}
\begin{array}{l}
E(1,M;\lambda D(1,M))=\ _0F_M([\; \;],[\ \underbrace{1,1,...1}_{M\ times}\ ],\lambda D(1,M))=\\
={\displaystyle\ :e^{-a^\dag a}\sum_{l=0}^\infty\frac{(a^\dag a)^l}{l!}\ _MF_M([\ \underbrace{1+l,1+l,...,1+l}_{M\  times}\ ],[\underbrace{1,1,...,1}_{M\ times}],\lambda a):\ ,}
\end{array}
\end{equation}
which indicates a pattern appearing in the course of this procedure.

Indeed, by evaluating the coherent state expectation value of Eq.(\ref{a1}) between $\langle z=1|\dots|z=1\rangle$ in the spirit of Eq.(\ref{Bell}) we obtain the hypergeometric generating functions of the numbers $B^{(M)}_{1}(n)$ as then
\begin{equation}
\label{a2}
\begin{array}{l}
{\displaystyle e^{-1}\sum\limits_{l=0}^{\infty}\frac{1}{l!}\ _MF_M([\underbrace{l+1, l+1, \dots, l+1}_{M\ times}],[\underbrace{1,1,...1}_{M\ times}],\lambda)=}\\
= {\displaystyle \sum_{n=0}^\infty B^{(M)}_{1}(n)\frac{\lambda^n}{(n!)^{M+1}}, \ \ \ \ \ M=1,2,\dots\ .}
\end{array}
\end{equation}
The hypergeometric generating function  of $B_{r}^{(M)}(n, x)$ for arbitrary $r$ and $M$  can also be obtained from Eq.(\ref{Dob}), and reads:
\begin{equation}
\label{a3}
\begin{array}{l}
{\displaystyle e^{-x}\sum\limits_{l=0}^{\infty}\frac{x^l}{l!}\ _MF_M([\underbrace{l/r+1, l/r+1, \dots, l/r+1}_{M\ times}],[\underbrace{1,1,...1}_{M\ times}],r^M\lambda)=}\\
{\displaystyle= \sum_{n=0}^\infty B^{(M)}_{r}(n,x)\frac{\lambda^n}{(n!)^{M+1}}\ .}
\end{array}
\end{equation}
In spite of their apparent complexity the l.h.s of the above equations can be straightforwardly handled by computer algebra systems \cite{Maple}.

\section{Conclusions and outlook}
\label{Conclusions}
We have found exact analytical expressions for the generalized Stirling numbers and generalized Bell polynomials which appear in the normal ordering of powers of  Laguerre-type derivative operators,   and have  provided a complete set  of hypergeometric generating functions for these quantities. The combinatorial aspect of the problem was demonstrated  by finding an exact mapping between the normal ordering and an enumeration of increasingly labelled, multivertex forests constructed according to a two-parameter $(r, M)$ prescription.  In this way analytical, numerical and combinatorial facets of this problem have been given a very complete treatment. We have  also used generalized Dobi\'nski relations to investigate the properties of these Laguerre-type differential operators. We provided a large number of operational formulas involving functions of Laguerre derivatives, which can alternatively be applied using the boson language. The framework developed above enables one to construct and analyze new coherent states relevant to nonlinear quantum optics, which will be  the subject of forthcoming research.

\section{Acknowledgments}

We wish to acknowledge support from Agence Nationale de la Recherche (Paris, France) under programme no. ANR-08-BLAN-0243-2.
Two of us, P.B. and A.H., wish to acknowledge support from the Polish Ministry of Science and Higher Education under grants no. N202 061434 and 202 107 32/2832.

\section{Appendix: Sheffer-type operators}\label{Appendix}

We derive Eq.(\ref{shef}) with the help of methods developed in Ref.\cite{PLA1}.
First, observe that $D(r,1)=a^\dag a^{r+1}+r a^r$ from which it follows that
$D^\dag(r,1)=(a^\dag)^{r+1} a +r (a^\dag)^r$ is an operator of Sheffer-type:
$D^\dag(r,1)=v(a^\dag)+q(a^\dag)a$ with $q(x)=x^{r+1}$ and $v(x)=rx^r$.
The normally ordered form of $\exp(\lambda D^\dag(r,1))$
is obtained by solving the linear differential equations (Eqs.(2) and (3) of Ref.\cite{PLA1})
for $T(\lambda,x)$ and $g(\lambda,x)$ yielding
\begin{eqnarray}
T(\lambda,a^\dag)=\frac{a^\dag}{(1-\lambda r (a^\dag)^r)^{1/r}},{\hskip 207pt{(A1)}}\nonumber
\end{eqnarray}
and
\begin{eqnarray}
g(\lambda,a^\dag)=\frac{1}{1-\lambda r (a^\dag)^r}\ .{\hskip 230pt{(A2)}}\nonumber
\end{eqnarray}
According to Eq.(29) of \cite{PLA1} the normally ordered form of $e^{\lambda D(r,1)}$ is
\begin{eqnarray}
e^{\lambda D(r,1)}=\left[e^{\lambda D^\dag(r,1)}\right]^\dag=\ :g(\lambda,a)e^{a^\dag(T(\lambda,a)-a)}:{\hskip 121pt{(A3)}}\nonumber
\end{eqnarray}
which gives Eq.(\ref{shef}).

\end{document}